\documentclass[aps,prb,twocolumn,preprintnumbers,amsmath,amssymb,superscriptaddress,shortbibliography]{revtex4-2}
\usepackage{graphicx}
\usepackage{epsfig}
\usepackage{dcolumn}
\usepackage{bm}
\usepackage{ulem}
\usepackage{color}
\usepackage{float}
\usepackage[colorlinks=true,linkcolor=blue,citecolor=blue,urlcolor=blue]{hyperref}%
\usepackage{verbatim}
\hypersetup{citecolor = blue}
\usepackage{subfigure}
\usepackage[ruled,vlined]{algorithm2e}

\def\avg#1{\left\langle#1\right\rangle}

\def\Re{{\rm Re}}

\def\trace#1{{\rm Tr}\left[#1\right]}

\def\be{\begin{equation}}       \def\ee{\end{equation}}
\def\bea{\begin{eqnarray}}      \def\eea{\end{eqnarray}}
\def\ba{\begin{array}}
\def\ea{\end{array}}
\def\bnum{\begin{enumerate} }
\def\enum{\end{enumerate}}

\def\=>{\Rightarrow}
\def\>{\rightarrow}

\def\eye2{Fathbb{I}}

\usepackage{ listings }

\def\Eq#1{Eq.~(\ref{#1})}
\def\Fig#1{Fig.~\ref{#1}}

\def\tr{\mathrm{tr}}
\def\Tr{\mathrm{Tr}}

\renewcommand{\c}[1]{{\cal #1}}

\renewcommand{\>}{\rangle}

\renewcommand{\Re}{{\rm Re}}
\newcommand{\p}{\partial}

\newcommand{\eq}[2]{
	\begin{equation}
	#1 \label{#2}
	\end{equation}
}

\newcommand{\mi}{\mathrm{i}}

\renewcommand{\rm}[1]{\mathrm{#1}}

\newcommand{\vect}[1]{\boldsymbol{#1}}

\usepackage{listings}
\usepackage{color}
\definecolor{lightgray}{gray}{1}

\begin{document}

\title{Mitigating the fermion sign problem by automatic differentiation}
\author{Zhou-Quan Wan}
\affiliation{Institute for Advanced Study, Tsinghua University, Beijing 100084, China}
\author{Shi-Xin Zhang}
\affiliation{Institute for Advanced Study, Tsinghua University, Beijing 100084, China}
\author{Hong Yao}
\email{yaohong@tsinghua.edu.cn}
\affiliation{Institute for Advanced Study, Tsinghua University, Beijing 100084, China}
\affiliation{State Key Laboratory of Low Dimensional Quantum Physics, Tsinghua University, Beijing 100084, China}

\begin{abstract}
As an \textit{intrinsically unbiased} method, the quantum Monte Carlo (QMC) method is of unique importance in simulating interacting quantum systems.
Although the QMC method often suffers from the notorious sign problem,
the sign problem of quantum models may be mitigated by finding better choices of the simulation scheme.
However, a general framework for identifying optimal QMC schemes has been lacking.
Here, we propose a general framework using automatic differentiation to automatically search for the best
QMC scheme within a given ansatz of the Hubbard-Stratonovich transformation,
which we call ``automatic differentiable sign optimization'' (ADSO).
We apply the ADSO framework to the honeycomb lattice Hubbard model with Rashba spin-orbit coupling and demonstrate that ADSO is remarkably effective in mitigating and even solving its sign problem. Specifically, ADSO finds a sign-free point in the model which was previously thought to be sign-problematic.
For the sign-free model discovered by ADSO, its ground state is shown by sign-free QMC simulations to possess spiral magnetic ordering; we also obtained the critical exponents characterizing the magnetic quantum phase transition.
\end{abstract}

\date{\today}
\maketitle

{\it Introduction.} The numerical study of quantum systems is of vital importance, especially in the context of strongly correlated systems which are in general analytically intractable in more than one dimension.
Due to their exponentially growing Hilbert space, numeric methods such as exact diagonalization usually fail when the system size is moderately large.
The quantum Monte Carlo (QMC) method can putatively overcome such an ``exponential wall'' by sampling a fraction of the Hilbert space stochastically.
The QMC method is {\it intrinsically unbiased}, making it one of the most powerful and successful methods to simulate quantum systems.
Unfortunately, the QMC method is often plagued by the notorious sign problem when dealing with fermion systems or frustrated spin models \cite{Hirsch1985,Takasu1986,Hatano1992}.
When the sign problem occurs, the simulation uncertainty increases exponentially with the system size and inverse temperature, rendering it infeasible in studying systems at low temperature or with large size \cite{Loh1990,Batrouni1990,RDevelopmentCoreTeam2011,Jia2014,Iglovikov2015, Kung2016,Huang2017}.
It has been desired for decades to solve the sign problem of interacting quantum models.

Tremendous progress has been made to solve the sign problem by identifying sign-free QMC schemes for quantum models with certain symmetries \cite{Wu2005,Wang2015,Li2015,Li2016,Wei2016} (see, e.g., Ref. \cite{Li2019} for a recent review). In studying these fermion models by the sign-problem-free QMC method, fruitful physics has been revealed  (see, e.g., Refs. \cite{Capponi2001,Assaad2005,Hohenadler2011,Nandini2011NP, Berg2012,Cai2013,Wang2014a,Li2015a,Schattner2016,Berg2016PRL, Li2016a,Assaad2016,He2016,Trebst2016,Li2017,Gazit2017,Li2017b, Li2017a,Bercx2017,Qin2017,Li2018a,Esterlis2018, Gazit2018PNAS,Esterlis2019,Zhang2019a,Xu2019,Lang2019, Li2019a,ZXLi2019,Berg2019,Costa2020,Sato2020,Xu2020, Berg2020PRR,Otsuka2016,Liu2019,Liu2021}).
Nonetheless, \textit{generically} solving the sign problem of quantum models is almost impossible as it has been proved that the sign problem complexity is NP-hard \cite{Troyer2005}. Moreover, it was shown recently that interacting models whose ground states feature certain properties such as a gravitational anomaly may have an intrinsic sign problem \cite{Hastings2016,Ringel2017,Golan2020,Smith2020}.
Fortunately, for a given \textit{specific} quantum model it is still possible to solve or mitigate its sign problem.
Efforts along this direction have been made recently;
sign problem mitigation was studied using basis transformation \cite{Shinaoka2015,levy2019,torlai2019,hangleiter2019,Klassen2019,Marvian2018,Kim2020}, Lefschetz thimbles \cite{Ulybyshev2019,Ulybyshev2020,Alexandru2020}, and machine learning techniques \cite{Broecker2017,LiangFu2017,Wynen2020}.
However, a universal framework for solving or mitigating the sign problem is still lacking.

Here, we fill in this gap by constructing a general framework of sign optimization in the determinant quantum Monte Carlo (DQMC) method. The DQMC method was introduced by Blankenbecler, Scalapino, and Sugar (BSS) \cite{Blankenbecler1981} and has been extensively used in simulating interacting fermion models.
Note that the severity of the sign problem in the DQMC method crucially depends on the scheme of Hubbard-Stratonovich (HS) transformation.
Different forms of HS transformations were proposed in the early stages of developing the DQMC method
\cite{Hirsch1983, Hirsch1986, Buendia1986,Batrouni1990,Batrouni1993,LIANGCHEN;A.-M.S.TREMBLAY1992}.
Nonetheless, previous HS transformations employed in simulations are quite limited in form and are constrained to \textit{no} spatial dependence.
It is desired to construct sufficiently general HS transformations and then find the optimized one for the sign of a given model.
In this Research Letter, we propose a general framework to realize sign optimization by parametrizing HS transforms continuously and optimizing the sign using automatic differentiation (AD) \cite{Bartholomew-Biggs2000,Gune2018,Margossian2019}.
We call it ``automatic differentiable sign optimization'' (ADSO). (AD is a powerful method for optimization that is widely encountered in machine learning and features various applications in computational physics \cite{Hubig2019,Liu2020,Hasik2020,Coopmans2020,Pakrouski2019,Chen2020, Sorella2010,Xie2020,Liao2019,Zhang2019}.)
ADSO is a general framework for mitigating the sign problem, applicable to most quantum lattice fermion models. We believe that ADSO will shed light on the nature of sign problem.

We further demonstrate the effectiveness of the general ADSO framework by applying it to the Rashba-Hubbard model (the usual Hubbard model plus Rashba couplings) on a honeycomb lattice. Although the Rashba-Hubbard model at half filling was known to be sign-problematic \cite{Hohenadler2012}, we show that its sign problem can be significantly mitigated by ADSO, which leads power-law acceleration. More remarkably, with the assistance of ADSO, we find a sign-free point in the model. This leads to an exponential acceleration in simulations and allows one to reliably obtain its physical properties by the sign-free QMC method. For the sign-free model identified by ADSO, its ground state is shown by large-scale QMC simulations to possess spiral magnetic ordering (as shown in \Fig{FSS} below). We further obtained critical exponents characterizing the quantum phase transition between the Dirac semimetal at weak Hubbard interaction and the spiral magnetic ordered state at strong interaction.

{\it The DQMC method and the sign problem.}
The DQMC method is widely used in simulating interacting fermion models.
To study equilibrium properties of an interacting fermion model described by Hamiltonian $\hat H=\hat H_0 +\hat H_I$ with $\hat H_0$ being the non-interacting term and $\hat H_I$ being the quartic or interacting term, one normally computes the expectation value of some observable $\hat O$: $
\langle\hat O\rangle=\frac{\Tr{(\hat O e^{-\beta \hat  H})}}{\Tr{(e^{-\beta \hat H})}}$,
where $\beta\!=\!1/T$ is the inverse temperature.
Using the Suzuki-Trotter decomposition \cite{Trotter1959,Suzuki1976} along the imaginary time direction, we obtain the density matrix
$e^{-\beta \hat H} = \prod_{l=0}^{L-1} e^{-\Delta \tau \hat H} \simeq\prod_{l=0}^{L-1}  e^{-\Delta \tau \hat H_0}e^ {-\Delta \tau \hat H_I}$,
where $\beta=L\Delta\tau$.
To deal with the quartic term $\hat H_I$, one can convert it into quadratic forms by performing HS transformations; the price to pay is the introduction of auxiliary fields.
A general form of HS transformation is given by
\bea\label{HS}
e^{-\Delta \tau \hat H_I} = \sum_{s} \eta(s)e^{\hat V(s)},
\eea
where $s$ represents auxiliary fields, $\hat V(s)=c^\dagger V(s) c$ are quadratic fermion operators with the matrix $V(s)$ and fermion creation operators $c^\dag$ (indices in $c^\dag$ are implicitly included), and $\eta(s)$ is a prefactor. For simplicity we assume that $s$ take discrete values, though continuously-valued auxiliary fields \cite{Beyl2018} can also be treated in ADSO.
With HS transformation at every time slice $l$,
we obtain the HS decoupled form of the density matrix:
$e^{-\beta \hat H} = \sum_{\vect s}\prod_{l=0}^{L-1}\eta (s_l)e^{-\Delta\tau \hat H_0 }e^{\hat V(s_l)}\!=\!\sum_{\vect s} \hat \rho_{\vect s}$,
where $\vect s\!=\!\{s_{l}\}$ represent an auxiliary-field configuration.

Then, the expectation value of observable $\hat O$ is given by
$\langle \hat O \rangle
=\frac{\sum_{\vect s} w(\vect s)  O(\vect s)}{\sum_{\vect s} w(\vect s)},
$
where $O(\vect s)$ is the expectation of $\hat O$ in the auxiliary-field configuration $\vect s$ and $w(\vect s)= \Tr(\hat \rho(\vect s))=\eta (\vect s)\det\left(\mathbb{I}+\prod_{l=0}^{L-1}e^Ke^{V(s_l)}\right)$ is the Boltzmann weight of auxiliary-field configuration $\vect s$ with $K$ being the matrix obtained from $-\Delta \tau \hat H_0=c^\dag K c$ and $\eta(\vect s)=\prod_{l= 0}^{L- 1}\eta(s_l)$.
To obtain $\langle \hat O \rangle$ by the QMC method, one computes the expectation of $O(\vect s)$ with $\vect s$ sampled from an unnormalized distribution $w(\vect s)$, namely $\langle\hat O\rangle=\avg{O(\vect s)}_{\vect s\sim w(\vect s)}$.
However, there is no guarantee that $w( \vect s)$ is always positive. When $w(\vect s)$ can take both positive and negative (sometimes complex) values, we have the so-called sign problem.

When the sign problem appears, the absolute value of $w(\vect s)$ can be used to sample the configurations by absorbing the sign or phase factor $e^{\mi \varphi(\vect s)}=w(\vect s)/|w(\vect s)|$ into observables:
$\big\langle O(\vect s)\big\rangle_{\vect s\sim w(\vect s)}=\frac{\langle e^{ \mi \varphi(\vect s)}O(\vect s)\rangle_{\vect s\sim |w(\vect s)|}}{\langle e^{ \mi \varphi(\vect s)}\rangle_{\vect s\sim |w(\vect s)|}}$, where the denominator and numerator can be calculated stochastically using the Markov chain Monte Carlo method with the auxiliary fields sampled from the distribution $|w(\vect s)|$. The denominator is the so-called average sign $S$ in the QMC method:
$S\equiv \big\langle e^ {\mi \varphi(\vect{s})}\big\rangle_{\vect s\sim |w(\vect s)|}=\frac{\sum_{\vect s} w(\vect s)}{\sum_ {\vect s} |w(\vect s)|}$. As the partition function $Z=\Tr(e^{-\beta \hat H})=\sum_{\vect s} w(\vect s)$ is always positive, the average sign $S$ must be positive, and it can be easily proved that $0\!<\!S\!\leq\! 1$. It was observed \cite{Loh1990} that the average sign decays exponentially with system size $N$ and inverse temperature $\beta$ as
$S \sim e^{-\kappa N\beta}$ for sufficiently large $N$ and $\beta$,
where $\kappa$ is a constant. For the sign-problematic (sign-free) QMC method, $\kappa>0$ ($\kappa=0$). When the sign problem occurs, to obtain the value of $\langle \hat O\rangle$ within a given accuracy, the needed QMC simulation time $M$ increases exponentially with size and inverse temperature: $
M\sim \frac{1}{S^2}\sim e^{2\kappa N\beta}$.
This exponential complexity greatly hinders the feasibility of applying the QMC method to study interacting systems with large size or low temperature; reducing $\kappa$ means sign mitigation and {\it power-law} acceleration. When the sign problem is solved (namely, what we have is sign-free), $M$ is reduced to power-law complexity, $M\sim N^3\beta$; solving the sign problem represents {\it exponential} acceleration.

{\it The ADSO framework.} The average sign $S$ or the prefactor $\kappa$ discussed above is \textit{not} an intrinsic property of a quantum model; instead it crucially depends on how the HS transformation is performed in the DQMC method. For a given model, a smaller $\kappa$ implies less severe sign problem. In other words, mitigating the sign problem is equivalent to reducing $\kappa$ by identifying an optimal HS transformation.
Suppose we have a set of possible HS transformations that can be parametrized by continuous parameters $\vect \xi$; the form of the HS transformation in \Eq{HS} now becomes
\eq{
e^{-\Delta \tau \hat H_I} = \sum_{s} \eta(\vect \xi,s)e^{\hat V(\vect \xi, s)}=\sum_{s}\eta(\vect \xi, s)e^{c^\dagger V(\vect \xi, s) c}.
}{HS2}
Consequently, $w(\vect \xi,\vect s)=\eta (\vect\xi,\vect s)\det[\mathbb{I}+\prod_{l=0}^{L-1}e^Ke^{V(\vect\xi,s_l)}]$, $S(\vect \xi)$, and $\kappa(\vect \xi)$ can all depend on the HS parameters $\vect \xi$.
Sign mitigation becomes an optimization problem in the parameter space of $\vect{\xi}$.

Here, we choose $\ln S$ instead of $S$ as our objective function for optimization and would like to maximize $\ln S$ (equivalently maximizing $S$).
We do not use $S$ directly because it may lead to vanishingly small gradients due to the possible exponential smallness of $S$. Using the fact that the partition function $Z$ of a given model is independent of $\vect \xi$, we obtain the differentiation of $\ln S$ as
$d\ln S
=-\Re \avg{\frac{dw(\vect \xi, \vect s)}{w(\vect \xi, \vect s)}}_{\vect s\sim{|w(\vect \xi, \vect s)|}}$ (see the Supplemental Materials (SM) for details \footnote{See Supplemental Materials for details. The SM of this work includes: 1. A brief introduction to automatic differentiation. 2. The detailed algorithm for calculating gradients of Sign using AD. 3. Proof of sign-problem free points in Rashba-Hubbard model. 4. Complementary QMC results of the sign-free Rashba-Hubbard model at $\lambda_R/ t=\sqrt 2$. 5. Examples of parameterized HS transformations. 6. Complementary results for different $\lambda_R$}).
Note that sign averaging is not involved here, which means computing the gradients itself is actually {\it sign-free}.
It is interesting that gradients of $\ln S$ could be efficiently and reliably calculated even though it is difficult to compute $S$ accurately.
Remarkably, the ADSO framework itself is sign-free; thus the ADSO framework can be directly applied on large size systems of interest. See the SM for details \cite{Note1} of computing the differentiation $\frac{dw(\vect \xi, \vect s)}{w(\vect \xi, \vect s)}$ using AD. It turns out that only very limit computational resources in addition to the standard
DQMC algorithm are required in our ADSO framework.

Now we have all the ingredients to calculate the gradients.
It is worth noting that we shall collect the gradients of many samples similar to previous methods of combining AD with Monte Carlo sampling \cite{Willia1992,Kleijnen1996,Zhang2019}.
Stochastic gradient descent (SGD) is suitable in our case to optimize the target function $\ln S$ since the gradients are calculated in a stochastic way: $\vect \xi \to \vect \xi + \delta \vect \nabla_{\vect \xi} \ln S$, where $\delta$ is the learning rate.

{\it The honeycomb Rashba-Hubbard model.}
We now apply our general ADSO framework to the honeycomb lattice Hubbard model with Rashba spin-orbit couplings \cite{Bychkov1984a}. The Hamiltonian of the honeycomb Rashba-Hubbard model at half filling is given by
\begin{eqnarray}\label{rashabahamiltonian}
\hat H&=&-t\sum_{\avg{ij}} c^\dagger_{i\alpha} c_{j\alpha}+\lambda_R \sum_{\avg{ij}}\mi\hat z\cdot( \vect \sigma_{\alpha\beta}\times{\vect d}_{ij})c^{\dagger}_{i\alpha}c_{j\beta}\nonumber\\
&&+U\sum_i (n_{i\uparrow}-\frac{1}{2})(n_{i\downarrow}-\frac{1}{2}),
\end{eqnarray}
where $c^\dag_{i\alpha}$ creates an electron on site $i$  with spin polarization $\alpha=\uparrow,\downarrow$, $n_{i\alpha}=c^\dag_{i\alpha}c_{i\alpha}$, $\avg{ij}$ labels the nearest neighbor (NN) sites $i$ and $j$, $\vect{\sigma}$ represent Pauli matrices, and ${\vect d}_{ij}$ is the vector pointing from site $i$ to site $j$.
We set the hopping $t=1$ as the energy unit. $\lambda_R$ is the Rashba spin-orbit coupling and $U$ is the Hubbard interaction.
This model is relevant to single-layer graphene on a substrate or an interface; for instance, the Rashba spin-orbit coupling has been observed in a graphene interface \cite{Dedkov2008,Marchenko2012}.
The model is invariant under the particle-hole transformation $c_{i\sigma}\to (-1)^i \sigma c^\dag_{i\bar \sigma}$; it describes a system at half filling.
This model is known to be sign-free only when $\lambda_R=0$.
For any $\lambda_R>0$, this model was believed to be sign-problematic \cite{Hohenadler2012}.
A natural question to ask is what HS transformation can give rise to the most mitigated and even solved sign problem for $\lambda_R>0$.

For the repulsive Hubbard interaction, we consider a general HS transformation with the auxiliary fields on each site $i$ coupled to spin operators along the direction $\vect n_i= (\sin\theta_i\sin\phi_i,\sin\theta_i\cos\phi_i,\cos\theta_i)$ with two continuous parameters $\theta_i$ and $\phi_i$ \cite{LIANGCHEN;A.-M.S.TREMBLAY1992}:
\bea\label{generalhs}
e^{-\Delta\tau U (n_{i\uparrow}-\frac 1 2 )(n_{i\downarrow }-\frac 1 2 )}=\frac 1 2 e^{-U \Delta \tau /4}\sum_{s_i=\pm 1}e^{\lambda s_i c^\dag_i\vect\sigma\cdot \vect n_i c_i},~~~
\eea
where $\cosh \lambda\!=\!\exp(U\Delta\tau /2)$ and $s_i$ is the auxiliary field.
Since $s_i=\pm 1$, the HS parameters $\vect n_i$ feature the equivalence $\vect n_i \equiv -\vect n_i$; consequently, hereinafter we can assume $n_i^z\geq 0$ for any $i$.
For repulsive Hubbard interactions, uniform $\vect n_i=\hat z$ for all $i$ has been chosen conventionally. However, in trying to optimize for the best HS transformations, the ADSO framework will allow spatially nonuniform $\vect n_i$, which turns out to be crucial for mitigating or solving the sign problem of a model which was conventionally thought to be sign-problematic.

\begin{figure}[t]
\includegraphics[width=0.9\linewidth]{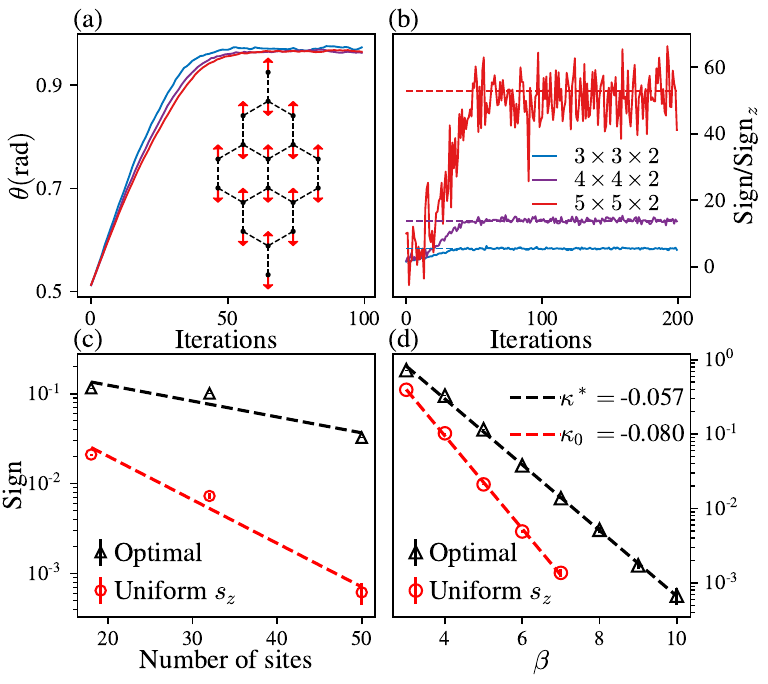}
\caption{Results of ADSO for the Rashba-Hubbard model on the honeycomb lattice with $\lambda_R=1.0$ and $U=6.0$. Here we fix $\phi_i=0$ and $\theta_{A,i}=-\theta_{B,i}=\theta$, which means $\theta$ is the only variational HS parameter. The corresponding pattern of $\vect n_i$ is shown in the inset of (a) (projected to the $xy$ plane).  Flow of HS parameters $\theta$ (a) and sign optimization results (b) for models with 3$\times$3$\times$2, 4$\times$4$\times$2, and 5$\times$5$\times$2 lattice sites (periodic boundary condition). In each iteration, the gradient is averaged using 336$\times$100 samples, where $336$ is the number of parallel Markov chains. The optimized values of parameters $\theta$ are nearly the same for different system sizes; (c) The scaling of sign average $S$ vs $N$;
(d) the scaling of sign average $S$ vs $\beta$;
} \label{lambda1}
\end{figure}

First, we apply ADSO to the Rashba-Hubbard model with $\lambda_R=1.0$ and $U=6.0$ to test the performance of the method.
For the $3\!\times\! 3\!\times\! 2$ lattice and starting from randomly chosen $\vect n_i$, we found that the optimized $\vect n_i$ is not uniform spatially, namely, $\vect n_i=(0,0.57,0.82)$ for the $i\in A$ sublattice and $\vect n_i=(0,-0.57,0.82)$ for the $i\in B$ sublattice as shown in the inset of \Fig{lambda1}(a).
Inspired by the optimal pattern obtained for the small system, we constrain the HS transformations to $(\theta_i,\phi_i)=(\theta,0)$ for the $i\in A$ sublattice and $(\theta_i,\phi_i)=(-\theta,0)$ for the $i\in B$ sublattice, where $0\leq \theta\leq \frac{\pi}{2}$ can vary to maximize the average sign. As shown in \Fig{lambda1}(a), we find that $\theta$ are converged to almost the same value for larger system sizes. This indicates that the optimized HS transformation does not change significantly with the system size; consequently, the optimized pattern obtained for relatively small system size can be directly used to perform QMC simulations on larger system size.

Moreover, as shown in \Fig{lambda1}(b), the larger the system size is, the more the sign problem improves. This indicates that the optimized HS transformation can reduce the prefactor $\kappa$ compared with the uniform $\vect n_i=\hat z$ scheme.
Since the Monte Carlo (MC) computation time $M$ scales as $M\sim \frac{1}{S^2}\sim e^{2\kappa N\beta}$, sign mitigation can be quantitatively characterized by how much the exponential prefactor $\kappa$ is reduced from optimizing HS transformations.
We use $\kappa^\ast$ ($S^\ast\sim e^{-\kappa^\ast N\beta}$ and $M^\ast\sim e^{2\kappa^\ast N\beta}$) to denote its value in the optimized HS transformation scheme and $\kappa_0$ ($S_0\sim e^{-\kappa_0 N\beta}$ and $M_0 \sim e^{2\kappa_0 N\beta}$) to denote the value in the spatially uniform HS scheme without optimization. Then, the computation is power-law accelerated from $M_0$ to $M^\ast\sim M_0^r$, where $r={\kappa^\ast/\kappa_0}$. As shown in \Fig{lambda1}(c) and (d), by comparing the scaling of the average sign $S$ versus $\beta$ and $N$, between the previously used HS scheme and the ADSO optimized one, we obtain $r=\kappa^\ast /\kappa_0\approx 0.7$.
The power-law acceleration with $r \approx 0.7$ can lead to tremendous acceleration especially when the system is large or the temperature is low.
For instance, for the lattice with $N=3$$\times$3$\times 2=18$ sites and inverse temperature $\beta=20.0$, the acceleration is already huge, and the computation is about $M_0/M^\ast$$\sim$$10^{7}$ times faster.

\begin{figure}[t]
\includegraphics[width=0.9\linewidth]{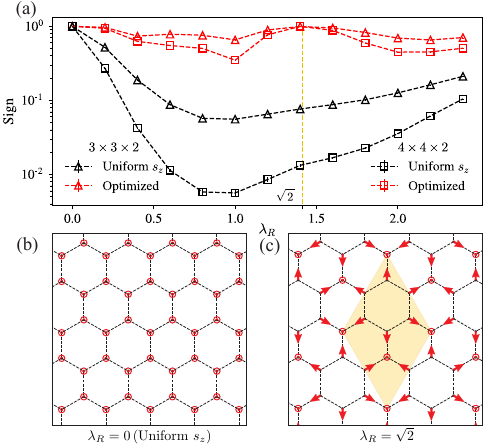}
\caption{Automatic sign optimization for the Rashba-Hubbard model with different parameters on the honeycomb lattice with $L=3,4$ (open boundary condition) and HS schemes for sign-free points. Here we choose $\beta=5,U=6$ and fix $t=1$.
(a) Optimized sign compared with the sign of the commonly used uniform $s_ z$ HS channel.
(b) Sign-problem free pattern of HS parameters $\vect n$ for the plain Hubbard model ($\lambda_R =0$), where red $\odot$ represent that $\vect n$ is pointing in the $\hat z$ direction (it is indeed the uniform $s_z$ channel).
(c) Sign-problem-free pattern of HS parameters $\vect n$ for the Rashba-Hubbard model at $\lambda_R/t=\sqrt 2$. Arrows represent the projection of $\vect n$ in the $xy$ plane. As indicated by the shaded region, $\vect n$ manifest a periodicity of $2\times 2$.
}
\label{sign_lambda}
\end{figure}

{\it The sign-free point identified by ADSO.} We further apply the ADSO method to the honeycomb Rashba-Hubbard model for various values of $\lambda_R$, as shown in \Fig{sign_lambda}(a).
It was previously known that the model is sign-free only for $\lambda_R=0$ (fixing $t=1$). For $\lambda_R=0$, the sign-free HS transformation is successfully found by ADSO, and it is indeed a uniform $\vect n_i=\hat z$ pattern, as shown in \Fig{sign_lambda}(b).
When $\lambda_R$ is increased from zero to finite values, the optimized sign is shown as in \Fig{sign_lambda}(a). Surprisingly, we notice that for $\lambda_R= 1.4$ the average sign has been optimized to $0.996$, which is very close to 1 (an average sign equal to 1 means that it is sign-free). The optimized sign being so close to 1 indicates that there may be an exactly sign-free point around this parameter region.
Indeed, we find that $\lambda_R=\sqrt 2$ is in fact an exactly sign-free point in the Rashba-Hubbard model using the HS transformation shown in \Fig{sign_lambda}(c) (see the SM for the exact proof \cite{Note1}) and this sign-free point was clearly indicated from the ADSO optimized sign being extremely close to 1. This successful example of solving the sign problem implies that ADSO has the potential possibility of helping people notice or identify new sign-free models.
\begin{figure}[t]
\includegraphics[width=0.9\linewidth]{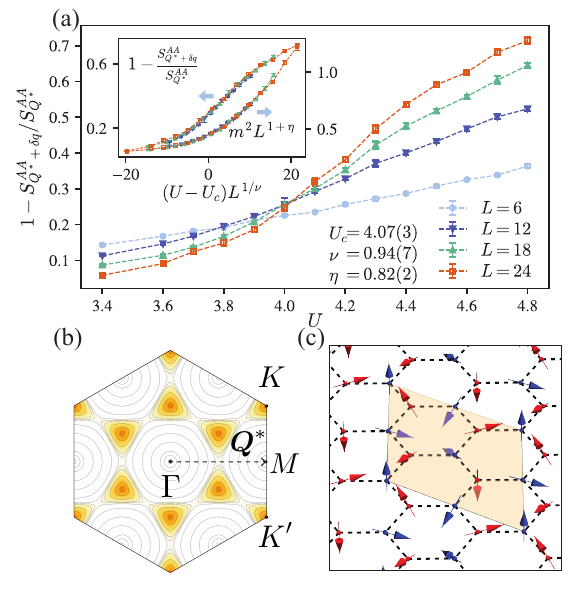}
\caption{The Rashba-Hubbard model at $\lambda_R/t=\sqrt 2$.
(a) Finite size scaling of correlation ratio $R=1-S^{AA}_{\vect{ Q^*}+\vect{ \delta q}}/S^{AA}_{\vect Q^*}$, where $S^{AA}_{\vect Q}$ is the spin structure factor defined as $\frac {1}{L^2}\sum_{\vect x_1,\vect x_2} e^{i \vect Q\cdot (\vect x_1-\vect x_2)}\left<\vect S^A(\vect x_1)\cdot \vect S^A(\vect x_2)\right>$, $\vect Q^*=\overline {\Gamma M}$ and $\vect{\delta q}=\vect a /L$ with $\vect a $ being the reciprocal lattice constant. Inset: Data collapse of $R$ and $m^2 \equiv S^{AA}_{\vect Q^*}/{L^2}$ using the critical value $U_c$ and the exponent $\nu,\eta$ extracted from the data of $L=12,18,24$ using the method in Refs.~\cite{Houdayer2004,Melchert2009}. Here we choose $\beta=L$ such that we can approach zero temperature in the thermodynamic limit.
(b) Contour plot of single particle gap of the Rashba-Hubbard model with $U=0$. It clearly shows eight two-component Dirac fermions in the Brillouin zone with two at the $K,K'$ point and six in the middle of $\Gamma-K$, $\Gamma-K'$.
(c) Visualization of magnetic order at $U>U_c$. This magnetic order manifests a periodicity of $2\times 2$ as shown by the shaded region. (This visualization is based on spin-spin correlations in different directions, see SM for details \cite{Note1}.)
}
\label{FSS}
\end{figure}

For the sign-free point $\lambda_R=\sqrt{2}$, we can perform large-scale QMC simulations to obtain reliably its quantum phase diagram as a function of $U$, as shown in \Fig{FSS}(a). For $0<U<U_c$, the ground state is a Dirac semimetal with \textit{eight} Dirac points (two-component Dirac fermion) as shown in \Fig{FSS}(b). For $U>U_c$, the ground state develops a spiral magnetic order as is shown in \Fig{FSS}(c). This phase transition should belong to the $N_f=16$ (using the convention in Ref.~\cite{Lang2019}) chiral Heisenberg Gross-Neveu-Yukawa (GNY) universality class \cite{Herbut2006}. From the finite-size scaling analysis of our QMC results, we obtain that the critical point is at $U_c= 4.07(3)$ with the correlation-length exponent $\nu=0.94(7)$ (correlation length $\xi\sim |U-U_c|^{-\nu}$) and order-parameter anomalous dimension $\eta=0.82(2)$. We highlight that these critical exponents of the $N_f=16$ chiral Heisenberg GNY universality class are obtained from sign-free QMC simulations (QMC results of critical exponents of the Heisenberg GNY universality class in 2+1D were obtained only with smaller $N_f$ \cite{Lang2019,Otsuka2016}).

As can be seen from the results above (both sign-mitigated and sign-solved cases), the optimized HS transformation, unlike the commonly used uniform $\vect n_i=\hat z$ decoupling scheme, is not spatially uniform. The optimal pattern of $\vect n_i$ can be different for different model parameters, which may be related to the properties of its underlying spin correlations of the ground states; for the two sign-free cases ($\lambda_R=0$ or $\lambda_R=\sqrt{2}$), the optimal patterns are indeed directly related to the magnetic ordering at strong $U$.

{\it Discussion and concluding remarks.}
The general framework of mitigating the sign problem in the DQMC method proposed in this Research Letter can be used in principle in any interacting quantum lattice models as long as its HS transformation can be continuously parametrized. For instance, by enlarging the auxiliary-field space or allowing hybrid decoupling schemes, further sign optimization may be obtained (see the SM for details \cite{Note1}). Moreover, the general idea of AD can be further applied to other types of QMC methods including world-line MC and hybrid MC whenever continuous parametrization can be implemented.

ADSO provides a general framework to mitigate the sign problem of interacting models; it worked remarkably well for the Rashba-Hubbard model which leads to power-law accelerations in general and even exponential acceleration for the sign-free point.
It is desirable to apply ADSO in the future to other strongly correlated models whose solutions remain elusive so far.
Moreover, ADSO has the potential possibility of identifying new sign-free models of interacting fermions.

{\it Acknowledgement.}
We thank Steve Kivelson and Zheng-Zhi Wu for helpful discussions and especially Zi-Xiang Li for related collaborations. This work is supported in part by the NSFC under Grant No. 11825404 (S.-X.Z., Z.-Q.W., and H.Y.), the MOSTC under Grant No. 2021YFA1400100 and No. 2018YFA0305604 (H.Y.), and the Strategic Priority Research Program of Chinese Academy of Sciences under Grant No. XDB28000000 (H.Y.).

\bibliographystyle{apsreve}
\bibliography{ADSO_arXiv}

\newpage
\begin{widetext}
\section*{Supplemental Materials}
\renewcommand{\theequation}{S\arabic{equation}}
\setcounter{equation}{0}
\renewcommand{\thefigure}{S\arabic{figure}}
\setcounter{figure}{0}
\subsection{A brief introduction to automatic differentiation}
Automatic differentiation (AD) is different from the conventional gradient evaluation methods including symbolic and numerical approaches.
AD can give gradients as accurate as symbolic differentiation while avoiding the difficulty of deriving the complex analytical expression.
By tracing the derivatives propagation of primitive operations via chain rules, numerically exact derivatives for almost all functions given by some programs can be achieved via AD. Here the program is specified by a computational graph composed of function primitives.

 \begin{figure}[htbp!]\centering
		\includegraphics[width=0.8\linewidth]{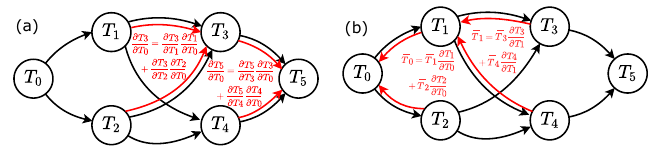}
		\caption{Forward mode (a) and reverse mode (b) automatic differentiation on computational graphs. Black arrows label the forward pass from inputs to outputs. Red arrows represent  forward chain rules in (a) and backpropagation for adjoints in (b).}  \label{ad}
\end{figure}

Based on the direction of tracing derivatives, there are two ways to compute the derivative on the graph with respect to the graph's inputs: the forward AD and backward AD. The forward/backward AD iteratively compute the recursive expression as shown in Fig.~\ref{ad}(a)/(b):
\bea
\begin{split}
\frac{\p T_i}{\p T_0}=\sum_{T_{i-1}\in \mathrm{parent} \{T_i\}}\frac{\p T_i}{\p T_{i-1}}\frac{\p T_{i-1}}{\p T_{0}}\ \ \ \ \text{ (forward)}\\
\overline{T}_i=\sum_{T_{i+1}\in \mathrm{child}\{T_{i}\}}\overline{T}_{i+1}\frac{\p T_{i+1}}{\p T_i}\ \ \ \ \text{ (backward)},
\end{split}
\label{adeq}\eea
where $T_i$ stands for nodes on the computational graph; $T_0$ is the input and $T_n$ the final output; $\overline{T}_i$ is called adjoint of $T_i$ defined as $\frac {\partial T_n}{\partial T_i}$. After going through the computational graph in the conventional direction, output of the function can be achieved (we call this a forward pass). In terms of forward AD, another forward pass will be conducted, and derivatives of all nodes with respect to the input node can be computed. While for backward AD, the computational graph will be further evaluated in the reverse direction (backward pass), and the gradients of final output $T_n$ with respect to all input nodes can be obtained. Clearly, backward AD is powerful when the number of input parameters is large (all gradients can be calculated in one backward pass).

In fact, the success of AD method is mainly due to the fact that almost all function primitives are automatic differentiable (the derivatives propagation can be expressed in close form). These function primitives are often implemented on top of AD infrastructure and the AD-aware primitives can be further customized for special purposes such as avoiding numerical instability in this paper.

\subsection{Calculating gradients of Sign using AD}
It is clear that the partition function $Z=\sum_{\vect s} w(\vect \xi,\vect s)$ of a quantum system is independent with parameters $\vect \xi$ in the HS transformation. As the average sign $S=\frac{\sum_{\vect s} w(\vect \xi,\vect s)}{\sum_{\vect s} |w(\vect \xi,\vect s)|}$, the differentiation of $\ln S$ can be evaluated as
\eq{
\begin{split}
d\ln S&=d\ln Z -d (\ln\sum_{\vect{s}} |w(\vect \xi, \vect s)|)= -\avg{\frac{d|w(\vect \xi, \vect s)|}{|w(\vect \xi, \vect s)|}}_{\vect s\sim{|w(\vect \xi, \vect s)|}}=-\Re \avg{\frac{dw(\vect \xi, \vect s)}{w(\vect \xi, \vect s)}}_{\vect s\sim{|w(\vect \xi, \vect s)|}},
 \end{split}
}{sm_dsign}
where the last equality is due to the fact that $\Re \frac{dw}{w}=\Re (\frac{d|w|}{|w|}+\mi d\varphi)=\frac{d|w|}{|w|}$. The reason why we don't use the sign $S$ itself as the target function becomes clearer using the fact that $d S=S \times  d\ln S$ will be very small if the sign problem is severe.

It seems that the differentiation $\frac{dw(\vect \xi, \vect s)}{w(\vect \xi, \vect s)}$ can be directly achieved using backward AD since the forward output $w(\vect \xi,\vect s)$ can be calculated as a determinant.
But it is actually trickier than that due to numerical instability of matrices product within determinants.
Since the forward evaluation of $w{(\vect{\xi}, \vect{s})}$ is plagued by lots of numerical stabilization procedures such as pivoted QR, it is hard to directly obtain the gradient via simple back propagation.
Furthermore, the gradient obtained in this way is not guaranteed to be numerical stable.
To address this problem, we further write the gradient as:
\bea\label{diff}
\frac{dw(\vect \xi, \vect s)}{w(\vect \xi, \vect s)}&&=\frac{d\eta(\vect \xi, \vect s)}{\eta(\vect \xi, \vect s)}+d\ln\det\left[\mathbb{I}+B(\vect \xi, \vect s)\right] \nonumber\\
&&=\frac{d\eta(\vect \xi, \vect s)}{\eta(\vect \xi, \vect s)}+\sum_{l=0}^{L-1}\trace{\overline G_l(\vect \xi, \vect s)B_{l}(\vect \xi, \vect s)^{-1}dB_l(\vect \xi, \vect s)},~~~
\eea
where $B(\vect \xi,\vect s)=\prod_{l=0}^{L-1} B_l(\vect \xi, s_l)$, $B_l(\vect \xi, s_l)=e^Ke^{V(\vect \xi, s_l)}$, and $\overline G_l=[\mathbb{I}+(B_{L-1}\cdots B_l)^{-1}(B_{l-1}\cdots B_0)^{-1}]^{-1}$.
The form of $\overline G_l$ is also encountered in usual DQMC when calculating equal-time Green's functions $G_l=(\mathbb I+B_{l-1}\cdots B_0 B_{L-1}\cdots B_l)^{-1}$.
As mentioned before, this kind of matrices product and inversion operation is not stable.
In this work, we use QR decomposition with column pivoting to stabilize the matrices product encountered in the calculation of gradients and equal-time green function.

\begin{algorithm}[H]
\SetAlgoLined
\label{QRPstable}
1) Compute pivoted QR: $B_0=QRP^T$\\
2) set $U_0=Q$, $D_0=\text{diag}(R)$, $V_0=D_0^{-1}RP^T$\\
3) \For {i in range(1,L)}{
\ \ \ \ \ \ Compute pivoted QR: $(B_iU_{i-1})D_{i-1}=QRP^T$\\
\ \ \ \ \ \ Set $U_i=Q$,$D_i=\text{diag}(R)$,$V_i=D_i^{-1}RP^TV_{i-1}$
}
4) Result: $B_{i}B_{i-1}\cdots B_0=U_iD_iV_i$
\caption{QRP stabilization}
\end{algorithm}

As shown in Alg.~(\ref{QRPstable}), matrices product can be decomposed into $UDV$, where $U$ is a unitary matrix, $D$ is a diagonal matrix and $V$ is supposed to be a well-conditioned matrix. Applying this algorithm, we get the decompositions:
\eq{
\begin{split}
B_{l-1}\cdots B_0&=U_RD_RV_R\\
B_{L-1}\cdots B_l&=V_LD_LU_L\\
\end{split}.
}{}

It is worth noting that the second decomposition is in a reverse order $VDU$ instead of $UDV$ which can be easily realized by processing the matrices from the left.
Using these results, equal-time Green's function and $\overline G_l$ can be calculated via numerical stable routines:
\eq{
\begin{split}
G_l&=(\mathbb I+ U_RD_RV_RV_LD_LU_L)^{-1}\\
&=U_L^{-1}((U_LU_R)^{-1}+D_RV_RV_LD_L)^{-1}U_R^{-1}\\
&=U_L^{-1}(D_L^b)^{-1}((D_R^b)^{-1}(U_LU_R)^{-1}(D_L^b)^{-1}+D_R^{s}V_RV_LD_L^ {s})^{-1}(D_R^b)^{-1}U_R^{-1},\\
\overline G_l&=(\mathbb I+ U_L^{-1}D_L^{-1}V_L^{-1}V_R^{-1}D_R^{-1}U_R^{-1})^{-1}\\
&=U_R(U_LU_R+D_L^{-1}V_L^{-1}V_R^{-1}D_R^{-1})^{-1}U_L\\
&=U_RD_R^s(D_L^sU_LU_RD_R^s+(D_L^b)^{-1}V_L^{-1}V_R^{-1}(D_R^b)^{-1})^{-1}D_L^sU_L,
\end{split}
}{stabilization2}

where $D_{L,R}=D_{L,R}^bD_{L,R}^s$ and
\eq{
\begin{split}
(D_{L,R}^b)_{ii} =
\begin{cases}
    (D_{L,R})_{ii} ,& \text{if } |(D_{L,R})_{ii}|> 1\\
    1,              & \text{otherwise}
\end{cases}\\
(D_{L,R}^s)_{ii} =
\begin{cases}
    1 ,& \text{if } |(D_{L,R})_{ii}|> 1\\
    (D_{L,R})_{ii},              & \text{otherwise}
\end{cases}
\end{split}.
}{dbs}
The rounding error caused by the addition in \Eq{diff} is eliminated by balancing the magnitude of the matrices to add up.

It is worth noting that these matrices $U_R,D_R,V_R,U_L,D_L,V_L$ are just by-products of the standard DQMC method since they are necessary ingredients to stabilize the calculation of equal-time green functions needed for updating the auxiliary field configurations. No extra computational resource apart from some memory costs is required to get these values.

After calculating these $\overline G_l$, we can use them to calculate the adjoint of $B_l$ which is defined as in \Eq{adeq}
that is
\eq{
\overline B_l \equiv \frac {\partial L}{\partial B_l}= -(\overline G_l B_l^{-1})^T,
}{adjointB}
where $L=\ln S$ is the target function.
Then we can send the adjoints of $B_l$ back into the computational graph of backward AD.
Then AD can do the remaining part of derivatives propagation.
Therefore, only very few computational resources in addition to standard DQMC algorithm are required in our ADSO framework.

\subsection{Proof of sign-problem free points in Rashba-Hubbard model}
\subsubsection{The usual sign-free model at $\lambda_R/t=0$}
This case is just the usual repulsive Hubbard model in a bipartite lattice at half filling, which is a prototype sign-free quantum model in DQMC. It is sign problem free using standard HHS transformation: $e^{-U \Delta\tau (n_{\uparrow}-\frac 1 2 )(n_{\downarrow }-\frac 1 2 )}=\frac 1 2 e^{-U \Delta \tau /4}\sum_{s=\pm 1}e^{\lambda s \hat \sigma_z}$.
The proof is as follows.
The weight is
\begin{equation}
w( \{\vect s\}) =\tr(\prod e^{\hat K} e^{\sum\lambda s_i \hat \sigma_i}).
\end{equation}
After applying a particle-hole transformation $c_{i\downarrow}\rightarrow (-1)^i c_{i\downarrow}^\dagger$, the weight changes into:
\begin{equation}
w( \{\vect s\}) =\tr(\prod e^{\hat K} e^{\sum\lambda s_i \hat n_i}) = \det(\mathbb I + \prod e^K e^{\sum \lambda s_i n_i})=w_ \uparrow(\{\vect s\})\times w_\downarrow(\{\vect s\}),
\end{equation}
where the matrix can be factorized into two identical blocks $w_\uparrow = w_ \downarrow \in \mathbb R$. Therefore, the model is sign-problem-free as $w= w_ \uparrow^2 \geq 0$.
Also notice that in this case, the system respects global $SU(2)$ symmetry.
Thus $e^{-U \Delta\tau (n_{\uparrow}-\frac 1 2 )(n_{\downarrow }-\frac 1 2 )}=\frac 1 2 e^{-U \Delta \tau /4}\sum_{s=\pm 1}e^{\lambda s \hat \sigma_{\theta,\phi}}$ is also a sign-problem-free HS transformation scheme as long as $\theta,\phi$ are identical in every site.

\subsubsection{The new sign-free model at $\lambda_R/t=\sqrt 2$}
For simplification, we set $\lambda_R/t=\tan \alpha$, 
where $\alpha =\arctan \sqrt 2$. We also introduce the spinor operator $\psi_{A/B,r}=(c_{A/B,r, \uparrow},c_{A/B,r, \downarrow})^T$. The kinetic term of Rashba-Hubbard model can be formulated as:
\eq{
\begin{split}
\hat H_{K} =H_{nn}+H_{Rashba}&= \sqrt{3}t\sum_{\{r\},i}\psi^\dagger_{A,r(i)}\exp\Bigg\{ -\mi \alpha \left(\begin{matrix}
0&-d_y^{(i)}-i d_x^{(i)}\\
-d_y^{(i)}+id_x^{(i)}&0
\end{matrix}\right)\Bigg\}
\psi_{B,r}+h.c.\\
&= \sqrt{3}t\sum_{\{r\},i}\psi^\dagger_{A,r(i)}\exp( -\mi \alpha (-d_y^{(i)}\sigma_x+d_x^{(i)}\sigma_y))
\psi_{B,r}+h.c..
\end{split}
}{rashaba2}
We consider gauged SU(2) transformation for this system $\psi_{A/ B, r}\rightarrow V_{A/B,r}\psi_{A/B, r}$, where $V_{A/B,r}$ are SU(2) matrices. Hubbard interaction is unchanged under SU(2) transformation while the kinetic term is changed into:
\eq{
\begin{split}
\hat H_{K} &= \sqrt{3}t\sum_{\{r\},i}\psi^\dagger_{A,r(i)} \left [V^\dagger_{A,r(i)}\exp\{ -\mi \alpha (-d_y^{(i)}\sigma_x+d_x^{(i)}\sigma_y)\}
V_{B, r}\right ] \psi_{B,r}+h.c..
\end{split}
}{rashaba3}
Note that if there exist a solution $V_{A/ B,r}$ so that $\left [V^\dagger_{A,r(i)}\exp( -\mi \alpha (-d_y^{(i)}\sigma_x+d_x^{(i)}\sigma_y))
V_{B, r}\right ]\propto \mathbb I$ for all $r$ and $i$, the model would be sign problem free since it can be transformed into the Hubbard model whose hopping is independent of spins via this SU(2) gauge transformation. 
The SU(2) gauge transformations conserve the SU(2) flux on any plaquette defined as $P = \prod_{(d_x,d_y)\in \text{plaquette}} \exp[ -\mi \alpha (-d_y\sigma_x+d_x\sigma_y)]$. 
For $\lambda_R/t=\sqrt 2$, this flux is $-\mathbb I$ for each hexagon plaquette. Consequently, it can be transformed into the Hubbard model with spin-independent hopping with $\pi$-flux on each plaquette which is sign problem free in the uniform spin HS channel. By transforming back to the origin model, we identify the sign-free HS transformation as $\frac 1 2 e^{-U \Delta \tau /4}\sum_s \exp(s\lambda\hat \sigma_{A/ B, r})$, where $\hat \sigma_{A/B,r}=V_{A/ B, r} \hat \sigma_z V_{A/ B, r}^\dagger$. Since $\pi$-flux model has global SU(2) symmetry, $\hat \sigma_z$ can be replaced by linear combination of Pauli operators $\hat \sigma_{\theta,\phi}$. Several sign-free HS transformations are shown in \Fig{sign_free_pattern}.

 \begin{figure}[htbp!]\centering
		\includegraphics[width=1.0\linewidth]{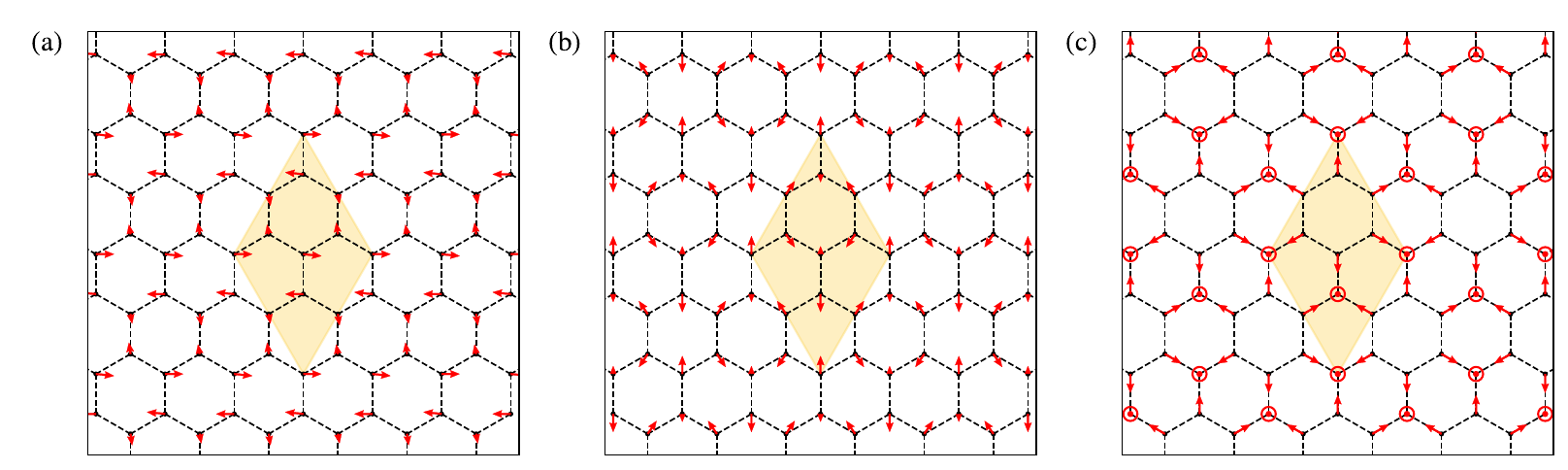}
		\caption{(a-c): Examples of sign-free HS parameters $\vect n_i$ in the case of $\lambda_R/ t= \sqrt 2$.  Arrows represent the projection of HS parameters $\vect n_i$ in $xy$-plane (we set $n_z>0$ using the equivalence $\vect n\equiv -\vect n$). These sign-free HS parameters $\vect n_i$ all manifest periodicity of $2\times2$ as indicated by shaded regions.}  \label{sign_free_pattern}
\end{figure}

With large enough interaction $U$, the usual Hubbard model and the $\pi$-flux Hubbard model on the honeycomb lattice all have an AF ground state. Since the Rashba-Hubbard model at these sign-problem-free points can be transformed into one of these two models, the ground states in these sign-problem-free cases are SDW generated by applying the gauged SU(2) transformation to the Neel AF order. It is worth noting that these magnetic orders are directly related to the sign-free HS transformation shown \Fig{sign_free_pattern}, which has $2\times 2$ periodicity. The phase transition of the usual Hubbard model with $0$-flux has been studied, which is shown to be in $N_f=8$ Gross-Neveu-Yukawa universality class with $U_c/t=3.85(2)$. In next section, we will present QMC results of this newly found sign-free model $\lambda_R/t=\sqrt 2$, whose magnetic quantum phase transition is characterized by a different universality class.

\subsection{QMC results of the sign-free Rashba-Hubbard model at $\lambda_R/t=\sqrt 2$}
Since the Rashba-Hubbard model at $\lambda_R/t=\sqrt 2$ is shown to be sign free, we can perform large-scale QMC simulations to investigate the phase diagram and phase transitions in this model. As shown in Fig. 3 of the main text, there are eight Dirac points in the Brillouin zone. In order to exactly access these eight Dirac points in the Brillouin zone, the system size must be an integer multiple of 6 (periodic boundary condition); thus we take system size $L=6,12,18,24$ in the simulations. We set $\Delta \tau =0.1$ in our simulations.

\begin{figure}[htbp!]\centering
	\includegraphics[width=1.0\linewidth]{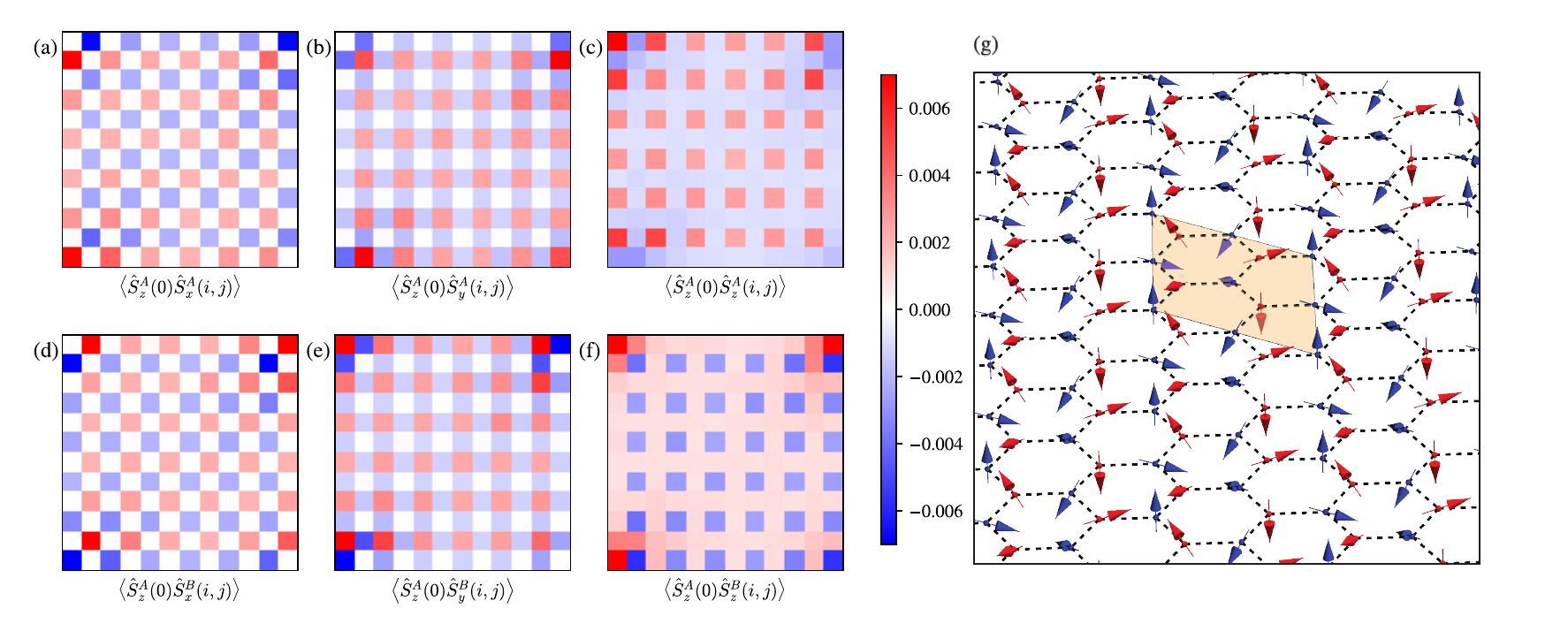}
	\caption{(a)-(f): Data of Spin-spin correlation in different directions. Here we use system size $L=12$ and $\beta=L=12$ with $U=4.5$. $i,j$ are indexes of lattice sites and $A,B$ are sublattice indexes. (g): Visualization of magnetic order using $\vect n_s(\vect x)\propto \left< \vect \hat S^A_z(0)\vect S^\alpha(\vect x)\right> $. The magnetic order manifests a periodicity of $2\times 2$ indicated by the shaded region (Fig. 3(c) shows a part of the figure here).}  \label{sscor}
\end{figure}

The magnetic order at large $U$ is characterized by spin-spin correlations $\left<\hat S_{a_1}^{\alpha_1}(\vect x_1) \hat S_{a_2}^{\alpha_2}(\vect x_2) \right> $ with $a_1,a_2\in\{x,y,z\}, \alpha_1,\alpha_2 \in \{A,B\}$ and also the spin structure factor $S^{\alpha_1,\alpha_2}_{\vect Q}$ defined as $ \frac {1}{L^2}\sum_{\vect x_1,\vect x_2} e^{i \vect Q\cdot (\vect x_1-\vect x_2)}\left<\vect S^{\alpha_1}(\vect x_1)\cdot \vect S^{\alpha_2}(\vect x_2)\right>$.
\Fig{sscor} present the results of spin-spin correlations for system size $L=12$ and $U=4.5$.
These correlations clearly manifest periodicity of $2\times 2$.
Detailed information of magnetic order can also be deduced from correlations in different directions.
Here we reconstruct the magnetic order using $\vect n_s(\vect x)\propto \left< \vect \hat S^A_z(0)\vect S^\alpha(\vect x)\right> $ as shown in \Fig{sscor}(g).

\begin{figure}[htbp!]\centering
		\includegraphics[width=1.0\linewidth]{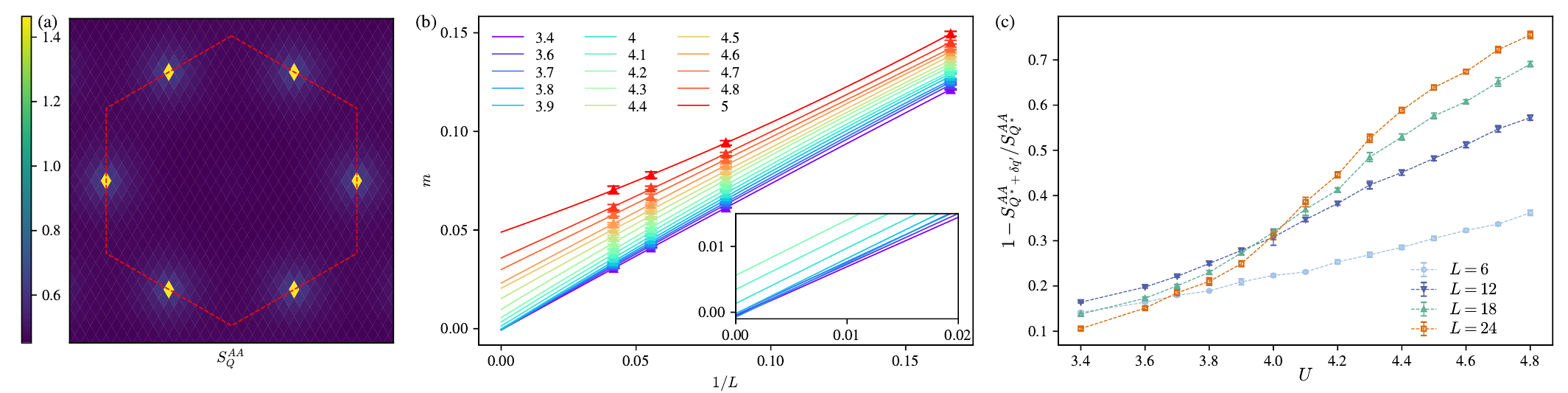}
		\caption{(a): Spin structure factor $S^{AA}_{\vect Q}$ for system size $L=24$ at $\beta=24, U=4.5$. It shows three peaks of in the first Brillouin zone (red dashed line). (b): Extrapolation of order parameter $m$ in thermodynamic limit (using quadratic function). (c): Spin correlation ratio $1-S^{AA}_{\vect Q^*+\delta \vect q'}/S^{AA}_{\vect Q^*}$ near the critical point. The crossing of curve indicates $U_c$ is around $4.0\sim 4.1$. This figure is similar to Fig.~3(a) but with different $\delta \vect q'=2\delta \vect q$.
		}  \label{FSS}
\end{figure}

\Fig{FSS}(a) shows the result of spin structure factor $S^{AA}_{\vect Q}$ in system size $L=24$ with $U=4.5$. It clearly shows 3 peaks in the Brillouin zone at $\vect Q^* = \overline {\Gamma M}$ and its $C_3$-symmetry related vectors. Thus, we can treat $m(L)=\sqrt {S^{AA}_{\vect Q^*}/L^2}$ as the order parameter.
By extrapolating $m(L)$ ($\beta = L$) in the thermodynamic limit, we find there is a phase transition near $U\simeq 3.9$ as shown in \Fig{FSS}(b). This is a preliminary estimation of the critical point since the number of system size is quite limited. Next, we perform a finite size scaling analysis to better estimate the quantum critical point. Near the vicinity of critical point, square of order parameter $m^2$ should obey the scaling law $m^2=L^{-1-\eta}\mathcal F (L^{1/\nu}(U-U_c),L/\beta)$ where $\mathcal F$ is a universal function, $\nu$ is the correlation length exponent and $\eta$ is the spin order parameter anomalous dimension. Here we assume the dynamic exponent $z=1$ since this quantum phase transition is expected to belong to the chiral Heisenberg Gross-Neveu-Yukawa (GNY) universality class. In the simulation, we fix $\beta=L$ to simplify the scaling function. 
The spin-spin correlation ratio defined as $R=1-\frac{S^{AA}_{\vect Q^*+\delta \vect q}}{S^{AA}_{\vect Q^*}}$ with $|\delta \vect q| \propto 1/L$ should obey the scaling law $R=\mathcal G(L^{1/\nu}(U-U_c))$ which is invariant at the critical point.
\Fig{FSS}(c) shows the finite-size scaling results of correlation ratio which is similar to Fig.~3(a) but with a different $\delta \vect q'=2\delta \vect q$.
The crossing of correlation ratio clearly indicates that $U_c$ is around $4.0\sim 4.1$. Then we collapse data of $m^2$ to extract the critical point $U_c$ and exponents $\nu,\eta$ as shown in Fig.~3(a). The $L=6$ data is excluded since the deviation from scaling law is too severe in this case due to finite size effect which can be easily seen from \Fig{FSS}(c). Finally, we obtain $U_c=4.07(3), \nu =0.94(7), \eta= 0.82(2)$.

\subsection{Continuously-parameterized HS transformations}
Continuous parametrization of HS transformations is essential to the ADSO framework. It is important to find a sufficiently general HS transformation which gives rise to reasonably good sign. We present a few parameterization approaches below where we use the Hubbard interaction as an example for most cases.

\subsubsection{Gauged HS transformation}

Gauged HS transformation was introduced in Ref.~\cite{LIANGCHEN;A.-M.S.TREMBLAY1992}. It was noticed that there exists some freedom in the conventional discrete HS transformations of the Hubbard interactions. For the repulsive case ($U>0$):
\eq{
e^{-U \Delta\tau (n_{\uparrow}-\frac 1 2 )(n_{\downarrow }-\frac 1 2 )}=\frac 1 2 e^{-U \Delta \tau /4}\sum_{s=\pm 1}e^{\lambda s c^\dag \vect \sigma \cdot \vect n c},
}{gaugedhs1}
where $c^\dag=(c^\dag_\uparrow,c^\dag_\downarrow)$ is a normal spinor, $\cosh \lambda =\exp(U\Delta \tau/2)$, and $\vect n=(\sin\theta\cos\phi,\sin\theta\sin\phi,\cos\theta)$. Here $\vect n$ or $(\theta,\phi)$ are the continuous parameters characterizing the HS transformation.  For the attractive case ($U<0$):
\eq{
e^{|U| \Delta\tau (n_{\uparrow}-\frac 1 2 )(n_{\downarrow }-\frac 1 2 )}=\frac{1}{2} e^{-|U| \Delta \tau /4}\sum_{s=\pm 1}e^{\lambda s \psi^\dagger {\vect \sigma} \cdot \vect n\psi},
}{gaugedhs2}
where $\psi^\dag=(c^\dag_\uparrow,c_\downarrow)$ is a Nambu spinor and $\cosh \lambda =\exp(|U|\Delta \tau/2)$. For $U<0$, the special case of $\vect n=\hat z$ is the familiar density (charge) decoupling scheme.

\subsubsection{Auxiliary fields with enlarged manifold}
When the manifold of auxiliary fields is larger than the minimal one, there is some freedom in choosing the value of parameters in the HS transformation. An interesting example for the Hubbard interaction was proposed by Hirsch [75] as follows:
\eq{
e^{-U\Delta\tau (n_{\uparrow}-\frac 1 2 )(n_{\downarrow }-\frac 1 2 )}=\frac{b}{2} e^{-U \Delta \tau /4} \sum_{s_\uparrow,s_\downarrow\in{\pm 1}}\exp(-\xi s_{\uparrow}s_{\downarrow}+\xi ' [s_{\uparrow}(2\hat n_{\uparrow}-1)+s_{\downarrow}(2\hat n_{\downarrow}-1)]),
}{}
where $\cosh(2\xi')=\frac{e^{U\Delta\tau /2 -e^{-2\xi }}}{1-e^{U\Delta\tau /2e^{-2\xi}}}$ and $b=\frac{1}{e^\xi +e^{-\xi}\cosh(2\xi')}$. Here $\xi$ is not fixed and can be treated as a continuous parameter. In general, an HS transformation can be continuously parameterized by extending the value space or manifold of auxiliary fields. For the Hubbard interaction, another continuous parameterization can be realized by extending the manifold from $\{\pm 1\}$ to $\{\pm n,\pm (n-1),\cdots,\pm 1,0\}$ as follows:
\eq{
e^{-U\Delta\tau (n_{\uparrow}-\frac 1 2 )(n_{\downarrow }-\frac 1 2 )}=\sum_{s=\{\pm n,\pm n-1, \cdots ,0\}}\eta(s) e^{\lambda(s) s\hat \sigma_z},
}{}
where $\eta(s)=\eta(-s)$ and $\lambda(s)=\lambda(-s)$ which satisfy
\eq{
\begin{split}
&\sum_{s=\{\pm n,\pm n-1, \cdots,0\}} \eta(s) = \exp(-U\Delta\tau/4)\\
&\sum_{s=\{\pm n,\pm n-1, \cdots,0\}} \eta(s)\cosh(\lambda(s)s)=\exp(U\Delta \tau/4).
\end{split}
}{}
There are totally $2(n+1)$ parameters in the HS parameters, including $(n+1)$ parameters $\eta(s)$ and $(n+1)$ parameters $\lambda(s)$. However, there are only 2 constraints. It is clear that this kind of HS transformations can be continuous parameterized by $2n$ parameters.

\subsubsection{Hybrid HS transformations}
When there are two or more different schemes of performing HS transformations for a certain type of interaction, one can introduce a hybrid HS transformation that can combine these schemes. For instance, suppose that there are two different HS schemes,
one can split $e^{-\Delta \tau \hat H_I}$ into two parts $e^{-\Delta \tau_1 \hat H_I}e^{-\Delta \tau_2\hat H_I}$, where $\Delta \tau_1+\Delta \tau_2=\Delta \tau$, and then perform different HS transformations in each part:
\eq{
e^{-\Delta \tau \hat H_I }=\sum_{s_1,s_2}\eta_1(s_1)\eta_2(s_2)e^{\hat V_1(s_1)}e^{\hat V_2(s_2)},
}{}
where $s_1,s_2$ are different auxiliary fields for the two different HS transformations.
We can use $0<\Delta \tau_1<\Delta\tau$ as a continuous parameter and thus the hybrid HS transformation can be continuously parameterized. This hybrid approach can also be combined with the former ways of extending HS transformations. Therefore the type of HS transformations for the better sign can be automatically selected by performing ADSO.

\subsection{More results for different $\lambda_R$}

\begin{figure}[ht]\centering
\includegraphics[width=1.0\linewidth]{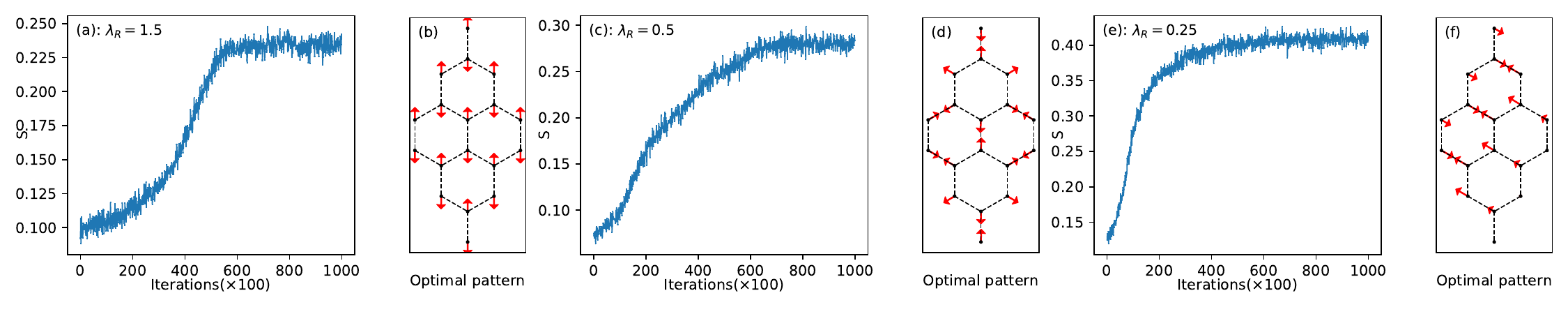}
\caption{Results of Hubbard-Rashba model on $3\times3$ honeycomb lattice with $\beta=5,U=6,t=1.0,\Delta \tau=0.1$ with periodic boundary condition. In each iteration, the gradients is average by $224$ samples, where $224$ is the number of paralleled Markov Chains.  (a),(c),(e): Optimization results for $\lambda_R=1.5,0.5,0.25$, respectively. (b),(d),(f): Optimal pattern of $\vect n$ for $\lambda_R=1.5,0.5,0.25$, respectively. Arrow represent the projection of $\vect n$ in $xy$ plane. Here we use the equivalence relation $\vect n\equiv -\vect n$ to make $\vect n_z>0$. The optimal pattern of $\lambda_R=1.5$ is still a AB sub-lattice pattern just like the case of $\lambda_R=1.0$, the optimal pattern of $\lambda_R=0.25$ is like a stripe pattern, and the optimal pattern of $\lambda_R=0.5$ has the periodicity of $\sqrt 3 \times \sqrt 3$.}  \label{additional1}
\end{figure}
\end{widetext}
\end{document}